\documentstyle[amstex,amssymb,12pt]{article}
\begin{document}

\author{{\bf Fabio Cardone}$^{a,b}${\bf , Alessio Marrani}$^{c,d}${\bf , Roberto
Mignani}$^{c,d}$ \\
%EndAName
$a$ Dipartimento di Fisica\\
Universit\`{a} dell'Aquila\\
Via Vetoio \\
67010 Coppito, L'Aquila, Italy\\
$b$ I.N.D.A.M. - G.N.F.M.\\
$c$ Dipartimento di Fisica ''E.Amaldi''\\
Universita' degli Studi ''Roma Tre''\\
Via della Vasca Navale, 84\\
I-00146 ROMA, Italy\\
$d$ I.N.F.N. - Sezione di Roma III\\
Via della Vasca Navale, 84\\
I-00146 ROMA, Italy}
\title{{\bf Boosts in an arbitrary direction and maximal causal velocities in a
deformed Minkowski space}}
\maketitle

\begin{abstract}
We discuss boosts in a deformed Minkowski space, i.e. a four-dimensional
space-time with metric coefficients depending on non-metric coordinates (in
particular on the energy). The general form of a boost in an arbitrary
direction is derived in the case of space anisotropy. Two maximal 3-vector
velocities are mathematically possible, an isotropic and an anisotropic one.
However, only the anisotropic velocity has physical meaning, being invariant
indeed under deformed boosts.
\end{abstract}

\section{Introduction\protect\bigskip}

In the last years, two of the present authors (F.C. and R.M.) proposed a
generalization\medskip\ of {\em Standard Special Relativity }(SR) based on
a\medskip\ ''deformation'' of space-time, assumed to be endowed with a
metric whose coefficients depend on the energy of\medskip\ the process
considered [1]. Such a formalism ({\em Deformed Special Relativity}, DSR)
applies in principle to {\em all\medskip } four interactions
(electromagnetic, weak, strong and gravitational) - at least as far as their
nonlocal behavior and\medskip\ nonpotential part is concerned - and provides
a metric representation of them (at\medskip\ least for the process and in
the\medskip\ energy range considered) ([1]-[5]). Moreover, it was shown that
such a formalism is actually a\medskip\ five-dimensional one, in
the\medskip\ sense that the deformed Minkowski space is embedded in\medskip\
a larger Riemannian manifold, with energy as fifth dimension [6].\medskip

In particular, the explicit expression of the Lorentz boost in the deformed
Minkowski space $\widetilde{M_{4}}$ for velocity along one of the coordinate
axes was derived in Ref.s [1] in the simpler case of an isotropic space. In
this paper, following the line of mathematical-formal research started with
[7], we want to derive the form of the deformed boost for velocity in an
arbitrary direction, and in the general case of spatial anisotropy. Such a
problem is far from being trivial, on account of the intrinsic anisotropic
nature of $\widetilde{M_{4}}$. Moreover, as we shall see, this will allow us
to further clarify the meaning of the maximal causal velocities in the
deformed space-time.

The organization of the paper is as follows. In Subsect. 2.1 we briefly
introduce the concept of deformed Minkowski space. Subsect. 2.2 deals with
the problem of the maximal velocity in $\widetilde{M_{4}}$, namely the
equivalent of the light speed in SR. It is shown that two different
mathematical procedures lead to two different maximal velocities, an
isotropic and an anisotropic one. Subsect.s 3.1 and 3.2 deal, respectively,
with a boost along one of the coordinate axis and in an arbitrary direction,
in the general spatially anisotropic case. The expression of the deformed
boots symmetric in space and time coordinates is given in Subsect. 2.3. We
derive in Subsect. 2.4 the generalized velocity composition law, which
allows us to state that only the anisotropic velocity is actually invariant
under deformed boosts, and has therefore a true physical meaning.

\bigskip

\section{\protect\bigskip Deformed Special Relativity in four dimensions
(DSR4)}

\subsection{Deformed Minkowski space-time}

The generalized (``deformed'') Minkowski space $\widetilde{M_{4}}$ (DMS4) is
defined as a space with the same local coordinates $x$ of $M_{4}$ (the
four-vectors of the usual Minkowski space), but with metric given by the
metric tensor\footnote{%
In the following, we shall employ the notation ''ESC on'' (''ESC off'') to
mean that the Einstein sum convention on repeated indices is (is not) used.}
\begin{gather}
\eta _{\mu \nu
}(x^{5})=diag(b_{0}^{2}(x^{5}),-b_{1}^{2}(x^{5}),-b_{2}^{2}(x^{5}),-b_{3}^{2}(x^{5}))=
\nonumber \\
\nonumber \\
\stackrel{\text{{\footnotesize ESC off}}}{=}\delta _{\mu \nu }\left[
b_{0}^{2}(x^{5})\delta _{\mu 0}-b_{1}^{2}(x^{5})\delta _{\mu
1}-b_{2}^{2}(x^{5})\delta _{\mu 2}-b_{3}^{2}(x^{5})\delta _{\mu 3}\right]
\end{gather}
where the $\left\{ b_{\mu }^{2}(x^{5})\right\} $ are dimensionless, real,
positive functions of the independent, non-metrical (n.m.) variable $x^{5}$
\footnote{%
Such a coordinate is to be interpreted as the energy $E$ (see Ref.s
[1]-[5]). However, \ since the metric coefficients $b_{\mu }^{2}(x^{5})$ are
{\em dimensionless}, their dependence on $E$ is actually of the type
\[
b_{\mu }^{2}\left( \frac{E}{E_{0}}\right) ,
\]
where $E_{0}$ plays the role of a {\em threshold energy}, characteristic of
the interaction considered (as before, see Ref.s [1]-[5]). Moreover, the
index $5$ explicitly refers to the above-mentioned fact that the deformed
Minkowski space can be {\em ''naturally'' embedded} in a five-dimensional
(Riemannian) space [6]. In this last framework, it is worth to give $x^{5}$
the dimension of a length.} . The generalized interval in $\widetilde{M_{4}}$
is therefore given by ($x^{\mu }=(x^{0},x^{1},x^{2},x^{3})=(ct,x,y,z)$, with
$c$ being the usual light speed in vacuum) (ESC on)
\begin{equation}
ds^{2}=b_{0}^{2}c^{2}\left( dt\right) ^{2}-\left[ b_{1}^{2}\left( dx\right)
^{2}+b_{2}^{2}\left( dy\right) ^{2}+b_{3}^{2}\left( dz\right) ^{2}\right]
=\eta _{\mu \nu }dx^{\mu }dx^{\nu }=dx\ast dx.
\end{equation}

The last step in (2) defines the scalar product $\ast $ in the deformed
Minkowski space $\widetilde{M_{4}}$ .\bigskip\ In order to emphasize the
dependence of DMS4 on the variable $x^{5}$, we shall sometimes use the
notation $\widetilde{M_{4}}(x^{5})$. It follows immediately that it can be
regarded as a particular case of a Riemann space with null curvature.

Let us stress that metric (1) is supposed to hold at a {\em local} (and not
global) scale. We shall therefore refer to it as a ``{\em topical}''
deformed metric, because it is supposed to be valid not everywhere, but only
in a suitable (local) space-time region (characteristic of both the system
and the interaction considered).

The two basic postulates of DSR4 (which generalize those of standard SR) are
[1]:

\ \ 1- {\em Space-time properties: } Space-time is homogeneous, but space is
not necessarily isotropic; a reference frame in which space-time is endowed
with such properties is called a ''{\em topical'' inertial reference frame}
(TIRF). Two TIRF's are in general moving uniformly with respect to each
other (i.e., as in SR, they are connected by a ''inertiality'' relation,
which defines an equivalence class of $\infty ^{3}$ TIRF );

2- {\em Generalized Principle of Relativity }(or {\em Principle of Metric
Invariance}): All physical measurements within each TIRF must be carried out
via the {\em same} metric.

The metric (1) is just a possible realization of the above postulates. We
refer the reader to Ref.s [1]-[5] for the explicit expressions of the
phenomenological energy-dependent metrics for the four fundamental
interactions.

In $\widetilde{M_{4}}$, it is possible {\it a priori }consider two scalar
products between trivectors {\footnotesize \ }$\underline{v_{1}}$%
{\footnotesize \ }, $\underline{v_{2}}$: the standard, Euclidean one $\cdot $
, defined by means of the metric tensor $g_{ik}=\delta _{ik}$ (here and in
the following small Latin indices will range in $\left\{ 1,2,3\right\} $),
and the deformed one, induced by the deformed scalar product $\ast $ in $%
\widetilde{M_{4}}$ , and defined by means of the metric tensor $-\eta
_{ik}(x^{5})\stackrel{\text{ESC off}}{=}b_{i}^{2}(x^{5})\delta _{ik}$%
{\footnotesize \ }(where the sign $-$ is obviously introduced in order to
get a positive trivector norm) as follows:
\begin{gather}
\underline{v_{1}}\ast \underline{v_{2}}\equiv -\sum_{i=1}^{3}\eta
_{ij}(x^{5})\left( v_{1}\right) ^{i}\left( v_{2}\right)
^{j}=\sum_{i=1}^{3}b_{i}^{2}(x^{5})\delta _{ij}\left( v_{1}\right)
^{i}\left( v_{2}\right) ^{j}=  \nonumber \\
\nonumber \\
=b_{1}^{2}(x^{5})\left( v_{1}\right) ^{1}\left( v_{2}\right)
^{1}+b_{2}^{2}(x^{5})\left( v_{1}\right) ^{2}\left( v_{2}\right)
^{2}+b_{3}^{2}(x^{5})\left( v_{1}\right) ^{3}\left( v_{2}\right) ^{3}
\end{gather}

Notice that, in the case of space isotropy ( $b_{i}(x^{5})=b(x^{5}),\forall
i=1,2,3$ ), the metric tensor reads:

\begin{gather}
\eta _{\mu \nu
,iso}(x^{5})=diag(b_{0}^{2}(x^{5}),-b^{2}(x^{5}),-b^{2}(x^{5}),-b^{2}(x^{5}))=
\nonumber \\
\nonumber \\
=\delta _{\mu \nu }\left[ \delta _{\mu 0}b_{0}^{2}(x^{5})-(\delta _{\mu
1}+\delta _{\mu 2}+\delta _{\mu 3})b^{2}(x^{5})\right]
\end{gather}

\bigskip It is easily seen that the two (space) scalar products $\cdot $ and
$\ast _{iso}$ are proportional :
\begin{equation}
\underline{a}\cdot \underline{d}=\sum_{i=1}^{3}a^{i}d^{i}=\frac{1}{%
b^{2}(x^{5})}\sum_{i=1}^{3}b^{2}(x^{5})a^{i}d^{i}=\frac{1}{b^{2}(x^{5})}%
\underline{a}\ast _{iso}\underline{d}
\end{equation}

\subsection{The trivector ''maximal causal velocity''(m.c.v.) in SR and in
DSR4}

As is well known, the maximal causal speed in $M_{4}$ is obtained by putting
$ds^{2}=0$, whence

\begin{equation}
ds^{2}=0\Leftrightarrow c^{2}dt^{2}-dx^{2}-dy^{2}-dz^{2}=0\Leftrightarrow
\frac{dx^{2}+dy^{2}+dz^{2}}{dt^{2}}=c^{2}
\end{equation}

Then one interprets $c$ as the maximal causal speed {\em along any direction
}of the (Euclidean) space $R^{3}$ (embedded in the pseudoeuclidean Minkowski
space-time $M_{4}$). Such an interpretation is obviously based on the
physical fact that $c$ coincides with the light speed in vacuum, and on the
isotropy of $R^{3}$ \footnote{%
Actually, it can be shown that the existence of an invariant, real quantity,
having the dimensions of the square of a speed follows from the Principle of
Relativity and the properties of homogeneity and isotropy of space-time. The
value of such a speed must be experimentally determined in the framework of
the total class $C_{T}$ of the physical phenomena considered (for instance, {%
it is obviously $\infty $ for Galilei's relativity, when only mechanics is
considered, and $c$ for Einstein's relativity, when also electromagnetism is
taken into account). See Ref.s} [1], and references therein.}. Therefore $c$
represents the value of any of the three components of the vector m.c.v. of
SR, $\underline{u}_{SR}$, namely:

\begin{equation}
\underline{u}_{SR}=c(\widehat{x},\widehat{y},\widehat{z})
\end{equation}

Then, $c^{2}$ is not, in general, a square modulus, but the square of any
component of $\underline{u}_{SR}$, whose square modulus (obviously with
respect to the Euclidean scalar product $\cdot $), is instead:

\begin{equation}
\left| \underline{u}_{SR}\right| ^{2}\equiv \sum_{i=1}^{3}\left(
u_{SR}^{i}\right) ^{2}=3c^{2}
\end{equation}
so that
\begin{equation}
u_{SR}^{i}=\frac{1}{\sqrt{3}}\left| \underline{u}_{SR}\right| \text{ \ \ \ \
\ \ \ \ }\forall i=1,2,3
\end{equation}

The above procedure must be suitably modified in the DSR4 case, due to the
space anisotropy of $\widetilde{M_{4}}$ .

Actually, in order to sort out a single component of the 3-vector m.c.v., in
a general 4-d special-relativistic theory (characterized by a diagonal
metric tensor $g_{\mu \nu }(\left\{ x\right\} _{n.m.})$ , where $\left\{
x\right\} _{n.m.}$ is a set of non-metrical variables), one has to exploit a
{\em ''directional separation''} (or {\em ''dimensional separation''})
method, which consists of the following three-step recipe (ESC off
throughout):

1- Set $ds^{2}$ equal to zero:

\begin{equation}
ds^{2}=0\Leftrightarrow g_{00}(\left\{ x\right\}
_{n.m.})c^{2}dt^{2}+\sum_{i=1}^{3}g_{ii}(\left\{ x\right\}
_{n.m.})(dx^{i})^{2}=0
\end{equation}

2- In order to find the i-th component $u^{i}(\left\{ x\right\} _{n.m.})$ of
the m.c.v., put $dx^{j}=0$ $(j\neq i)$ , thus getting

\begin{equation}
g_{00}(\left\{ x\right\} _{n.m.})c^{2}dt^{2}+g_{ii}(\left\{ x\right\}
_{n.m.})(dx^{i})^{2}=0
\end{equation}

3- Evidence on the lhs of (11) a quantity with physical dimensions $\frac{%
\left[ \text{space}\right] }{\left[ \text{time}\right] }=\left[ \text{%
{\small velocity}}\right] $ ; at this point, we have two different subcases:

I) One carries to the lhs of (11) $\frac{dx^{i}}{dt}$ (which amounts to
consider the 3-d Euclidean product $\cdot $), thus getting an {\em %
anisotropic} m.c.v.:

\begin{equation}
u^{i}(\left\{ x\right\} _{n.m.})\equiv \frac{dx^{i}}{dt}=\frac{\left(
g_{00}(\left\{ x\right\} _{n.m.})\right) ^{1/2}}{\left( -g_{ii}(\left\{
x\right\} _{n.m.})\right) ^{1/2}}c\text{ \ \ \ \ \ }\forall i=1,2,3.
\end{equation}

II) One carries to the lhs of (11) $\left( -g_{ii}(\left\{ x\right\}
_{n.m.})\right) ^{1/2}\frac{dx^{i}}{dt}$ (which amounts to consider the 3-d
deformed product $\ast $ defined by $-g_{ij}(\left\{ x\right\} _{n.m.})$ $%
=\delta _{ij}\left| g_{ii}(\left\{ x\right\} _{n.m.})\right| $ , thus
getting an {\em isotropic} m.c.v.:

\begin{equation}
u^{i}(\left\{ x\right\} _{n.m.})\equiv \left( -g_{ii}(\left\{ x\right\}
_{n.m.})\right) ^{1/2}\frac{dx^{i}}{dt}=\left( g_{00}(\left\{ x\right\}
_{n.m.})\right) ^{1/2}c\text{ \ \ \ \ \ }\forall i=1,2,3.
\end{equation}

The two subcases I and II differ essentially by the different way of
implementing the space anisotropy. In the former case, the anisotropy is
embedded in the definition of m.c.v.; in the latter one, in the scalar
product\footnote{%
Of course, the procedure of ''directional separation'' gives (in either
subcase) the same standard result when applied to SR. In fact:
\begin{gather*}
u_{SR}^{i}=\left( -g_{ii}\right) ^{1/2}\frac{dx^{i}}{dt}=\left(
g_{00}\right) ^{1/2}c\text{ \ }=\frac{dx^{i}}{dt}= \\
\text{ \ } \\
=\frac{\left( g_{00}\right) ^{1/2}}{\left( -g_{ii}\right) ^{1/2}}c=c\ \ \ \
\forall i=1,2,3
\end{gather*}
\par
{}} \footnote{%
Let us notice that the directionally separating procedure can be
consistently applied only to (special- or general-relativistic) metrics
which are fully {\em diagonal}. This is obviously due to the mixings between
different space directions which arise in the case of non-diagonal metrics.}.

Specializing the above equations to the DSR4 framework, we get therefore, in
the two subcases:

I)

\begin{equation}
u_{DSR4,I}^{i}(x^{5})\equiv u^{i}(x^{5})=c\frac{b_{0}(x^{5})}{b_{i}(x^{5})},
\end{equation}
\begin{gather}
\left| \underline{u_{DSR4,I}}(x^{5})\right| =\left( \sum_{i=1}^{3}\left(
u_{DSR4,I}^{i}(x^{5})\right) ^{2}\right) ^{1/2}=  \nonumber \\
\nonumber \\
=cb_{0}(x^{5})\left( \frac{1}{b_{1}^{2}(x^{5})}+\frac{1}{b_{2}^{2}(x^{5})}+%
\frac{1}{b_{3}^{2}(x^{5})}\right) ^{1/2}
\end{gather}
The vector $\underline{u}$ is the (spatially) anisotropic generalization of
the maximal causal speed introduced in the (spatially) isotropic case [1] ;

\bigskip

II)

\begin{equation}
u_{DSR4,II}^{i}(x^{5})\equiv w^{i}(x^{5})=cb_{0}(x^{5}),
\end{equation}
\begin{equation}
\left| \underline{u_{DSR4,II}}(x^{5})\right| _{\ast }=\left(
\sum_{i=1}^{3}b_{i}^{2}(x^{5})\left( u_{DSR4,II}^{i}(x^{5})\right)
^{2}\right) ^{1/2}=cb_{0}(x^{5})\left(
b_{1}^{2}(x^{5})+b_{2}^{2}(x^{5})+b_{3}^{2}(x^{5})\right) ^{1/2}
\end{equation}
whence

\begin{equation}
u_{DSR4,II}^{i}(x^{5})=\left(
b_{1}^{2}(x^{5})+b_{2}^{2}(x^{5})+b_{3}^{2}(x^{5})\right) ^{-1/2}\left|
\underline{u_{DSR4,II}}(x^{5})\right| _{\ast }
\end{equation}
i.e in this subcase (unlike the previous one, see Eq.s (14),(15)) one can
state a proportionality relation by an overall factor (even if dependent on
the metric coefficients) between $u_{DSR4,II}^{i}(x^{5})$ and $\left|
\underline{u_{DSR4,II}}(x^{5})\right| _{\ast }$.

We have therefore shown that the two different procedures of directional
separation lead to two {\em different} mathematical definitions of maximal
velocity, an isotropic ($\underline{w}$) and an anisotropic ($\underline{u}$%
) one. The choice between them must be done on a physical basis (see
Subsect. 3.4).

\bigskip

\section{Boosts in DSR4}

\subsection{Boost direction along $\widehat{x^{i}}$ ($i=1,2,3$)}

It follows from the principles 1), 2) of DSR4 that the transformations among
TIRF's, called ''Deformed Lorentz transformations'' (DLT), are those
(homogeneous) coordinate transformations which leave the metric tensor $\eta
_{\mu \nu }$ invariant. Since the deformed Minkowski space is a special case
of a Riemann space, we can state that the DLT's are the space-time
rotational component of the (maximal) Killing group of $\widetilde{M_{4}}%
(x^{5})$. Then, physical laws are to be covariant with respect to such
generalized transformations.

By following a procedure analogous to that used in SR to derive LT's (see
Appendix A), we get, for a boost along $\widehat{x^{i}}$ , $\forall i=1,2,3$%
) (ESC off throughout) :
\begin{equation}
\left\{
\begin{array}{ccc}
\left( x^{\prime }\right) ^{i}=\left( x^{i}\right) ^{\prime } & = &
\begin{array}{c}
\begin{array}{c}
\begin{array}{c}
\begin{array}{c}
\widetilde{\gamma }(x^{i}-v^{i}t)=\ \ \widetilde{\gamma }\left( x^{i}-%
\widetilde{\beta }\dfrac{b_{0}(x^{5})}{b_{i}(x^{5})}ct\right)
\end{array}
\end{array}
\end{array}
\end{array}
\\
\left( x^{\prime }\right) ^{k\neq i}=\left( x^{k\neq i}\right) ^{\prime } & =
&
\begin{array}{c}
\begin{array}{c}
\begin{array}{c}
\begin{array}{c}
x^{k\neq i}\smallskip \smallskip
\end{array}
\end{array}
\end{array}
\end{array}
\\
t^{\prime } & = &
\begin{array}{c}
\widetilde{\gamma }\left( t-\dfrac{v^{i}b_{i}^{2}(x^{5})}{%
c^{2}b_{0}^{2}(x^{5})}x^{i}\right) =\widetilde{\gamma }\left( t-\dfrac{%
\widetilde{\beta }^{2}}{v^{i}}x^{i}\right)
\end{array}
\end{array}
\right. ,
\end{equation}
where:

\begin{equation}
\widetilde{\beta }=\widetilde{\beta ^{i}}=\frac{v^{i}b_{i}(x^{5})}{%
cb_{0}(x^{5})}\ ;
\end{equation}
\begin{equation}
\widetilde{\gamma }=\left[ 1-\left( \widetilde{\beta ^{i}}\right) ^{2}\right]
^{-1/2}=\left[ 1-\left( \frac{v^{i}b_{i}(x^{5})}{cb_{0}(x^{5})}\right) ^{2}%
\right] ^{-1/2}
\end{equation}
Of course, in the non-relativistic limit $\lim_{c\rightarrow \infty }$ , $%
\widetilde{\beta }\rightarrow 0^{+}$ and $\widetilde{\gamma }\rightarrow
1^{+}$ , so that the deformed boosts reduce to the Galilean transformations.

\subsection{Boost in a generic direction}

In this case, the relative velocity is $\underline{v}=v^{1}\widehat{x}+v^{2}%
\widehat{y}+v^{3}\widehat{z}$ , and we have to suitably generalize
definitions (20), (21) as follows \footnote{%
Notice that $\widetilde{\underline{\beta }}:=\underline{\left( \frac{v}{u}%
\right) }\neq \frac{\underline{v}}{u}$ . This follows from the anisotropy of
the 3-vector $\underline{u}$ , and it is to be compared with the SR case,
where $\underline{\beta }:=\underline{\left( \frac{v}{c}\right) }=\frac{%
\underline{v}}{c}$ . In general, it is possible to state that
\[
\underline{\left( \frac{m}{n}\right) }=\frac{1}{n}\underline{m}%
\Leftrightarrow \underline{n}=n(\widehat{x},\widehat{y},\widehat{z})
\]
i.e iff $\underline{n}$ is a spatially isotropic trivector.}:

\begin{equation}
\widetilde{\underline{\beta }}\equiv \underline{\left( \frac{v}{u}\right) }%
=\left( \frac{v^{1}b_{1}(x^{5})}{cb_{0}(x^{5})},\frac{v^{2}b_{2}(x^{5})}{%
cb_{0}(x^{5})},\frac{v^{3}b_{3}(x^{5})}{cb_{0}(x^{5})}\right)
\end{equation}
\begin{equation}
\widetilde{\gamma }\equiv \left( 1-\left| \widetilde{\underline{\beta }}%
\right| ^{2}\right) ^{-1/2}
\end{equation}
where (cfr. Eq.(14))

\begin{equation}
\underline{u}=\left( c\frac{b_{0}(x^{5})}{b_{1}(x^{5})},c\frac{b_{0}(x^{5})}{%
b_{2}(x^{5})},c\frac{b_{0}(x^{5})}{b_{3}(x^{5})}\right)
\end{equation}

In order to derive the expression of the deformed boost in a generic
direction, it is possible to use the same method of the previous case (see
Appendix A). However, it is simpler to consider the notion of parallelism
between 3-vectors in $\widetilde{M_{4}}(x^{5})$ \footnote{%
The definitions of parallelism and orthogonality are to be meant in the
sense of the deformed 3-d. scalar product $\ast $ (see Eq. (3)).} and
decompose the space vector $\underline{x}$ in two components, $\underline{%
x_{\parallel }}$ and $\underline{x_{\perp }}$, parallel and orthogonal,
respectively, to the boost direction $\widehat{v}$
\begin{equation}
\underline{x}=\underline{x_{\parallel }}+\underline{x_{\perp }},
\end{equation}

\begin{gather}
\underline{x_{\parallel }}\equiv \widehat{v}(\widehat{v}\ast \underline{x})=%
\frac{\underline{v}}{\left| \underline{v}\right| _{\ast }^{2}}(\underline{v}%
\ast \underline{x})=\frac{\underline{v}}{\underline{v}\ast \underline{v}}(%
\underline{v}\ast \underline{x})=  \nonumber \\
\nonumber \\
=\frac{\sum_{i=1}^{3}b_{i}^{2}(x^{5})v^{i}x^{i}}{%
\sum_{i=1}^{3}b_{i}^{2}(x^{5})\left( v^{i}\right) ^{2}}\underline{v}%
\stackrel{\text{{\footnotesize in DSR4: }}\widetilde{\underline{\beta }}%
\equiv \underline{\left( \frac{v}{u}\right) }\neq \frac{\underline{v}}{u}}{%
\neq }\widehat{\widetilde{\beta }}(\widehat{\widetilde{\beta }}\ast
\underline{x})=  \nonumber \\
\nonumber \\
=\frac{\widetilde{\underline{\beta }}}{\left| \widetilde{\underline{\beta }}%
\right| _{\ast }^{2}}(\widetilde{\underline{\beta }}\ast \underline{x})=%
\frac{\widetilde{\underline{\beta }}}{\widetilde{\underline{\beta }}\ast
\widetilde{\underline{\beta }}}(\widetilde{\underline{\beta }}\ast
\underline{x})=\frac{\sum_{i=1}^{3}b_{i}^{2}(x^{5})\widetilde{\beta }%
^{i}x^{i}}{\sum_{i=1}^{3}b_{i}^{2}(x^{5})\left( \widetilde{\beta }%
^{i}\right) ^{2}}\underline{\widetilde{\beta }}
\end{gather}
(with $\left| {}\right| _{\ast }$ denoting the absolute value of a trivector
{\em with respect to the deformed scalar product }${\em \ast }$, whereas the
notation $\left| {}\right| $ will be used for the trivector norm with
respect to the standard product $\cdot $)

\begin{equation}
x_{\parallel }^{i}\equiv \frac{\sum_{k=1}^{3}b_{k}^{2}(x^{5})v^{k}x^{k}}{%
\sum_{k=1}^{3}b_{k}^{2}(x^{5})\left( v^{k}\right) ^{2}}v^{i}\stackrel{\text{%
{\footnotesize in DSR4: }}\widetilde{\underline{\beta }}\equiv \underline{%
\left( \frac{v}{u}\right) }\neq \frac{\underline{v}}{u}}{\neq }\frac{%
\sum_{k=1}^{3}b_{k}^{2}(x^{5})\widetilde{\beta }^{k}x^{k}}{%
\sum_{k=1}^{3}b_{k}^{2}(x^{5})\left( \widetilde{\beta }^{k}\right) ^{2}}%
\widetilde{\beta }^{i},
\end{equation}
\begin{gather}
\underline{x_{\perp }}\equiv \underline{x}-\underline{x_{\parallel }}=%
\underline{x}-\frac{\sum_{i=1}^{3}b_{i}^{2}(x^{5})v^{i}x^{i}}{%
\sum_{i=1}^{3}b_{i}^{2}(x^{5})\left( v^{i}\right) ^{2}}\underline{v}\neq
\nonumber \\
\nonumber \\
\stackrel{\text{{\footnotesize in DSR4: }}\widetilde{\underline{\beta }}%
\equiv \underline{\left( \frac{v}{u}\right) }\neq \frac{\underline{v}}{u}}{%
\neq }\underline{x}-\frac{\sum_{i=1}^{3}b_{i}^{2}(x^{5})\widetilde{\beta }%
^{i}x^{i}}{\sum_{i=1}^{3}b_{i}^{2}(x^{5})\left( \widetilde{\beta }%
^{i}\right) ^{2}}\underline{\widetilde{\beta }},
\end{gather}

\begin{gather}
x_{\perp }^{i}\equiv x^{i}-\frac{\sum_{k=1}^{3}b_{k}^{2}(x^{5})v^{k}x^{k}}{%
\sum_{k=1}^{3}b_{k}^{2}(x^{5})\left( v^{k}\right) ^{2}}v^{i}\neq  \nonumber
\\
\nonumber \\
\stackrel{\text{{\footnotesize in DSR4: }}\widetilde{\underline{\beta }}%
\equiv \underline{\left( \frac{v}{u}\right) }\neq \frac{\underline{v}}{u}}{%
\neq }x^{i}-\frac{\sum_{k=1}^{3}b_{k}^{2}(x^{5})\widetilde{\beta }^{k}x^{k}}{%
\sum_{k=1}^{3}b_{k}^{2}(x^{5})\left( \widetilde{\beta }^{k}\right) ^{2}}%
\widetilde{\beta }^{i}.
\end{gather}
It is easily checked that indeed

\begin{gather}
\underline{x}\ast \underline{v}=\sum_{i=1}^{3}b_{i}^{2}(x^{5})x^{i}v^{i}=%
\frac{\sum_{i=1}^{3}b_{i}^{2}(x^{5})x^{i}v^{i}}{%
\sum_{i=1}^{3}b_{i}^{2}(x^{5})\left( v^{i}\right) ^{2}}%
\sum_{k=1}^{3}b_{k}^{2}(x^{5})\left( v^{k}\right) ^{2}=  \nonumber \\
\nonumber \\
=\frac{\sum_{i=1}^{3}b_{i}^{2}(x^{5})x^{i}v^{i}}{%
\sum_{i=1}^{3}b_{i}^{2}(x^{5})\left( v^{i}\right) ^{2}}\underline{v}\ast
\underline{v}=\underline{x_{\parallel }}\ast \underline{v}=\left| \underline{%
x_{\parallel }}\right| _{\ast }\left| \underline{v}\right| _{\ast },
\end{gather}
\begin{equation}
\underline{x_{\perp }}\ast \underline{v}=\underline{x}\ast \underline{v}-%
\underline{x_{\parallel }}\ast \underline{v}=\underline{0}.
\end{equation}

We can now apply the boost (19) to $\underline{x_{\parallel }}$ and $%
\underline{x_{\perp }}$ , thus getting
\begin{equation}
\left\{
\begin{array}{ccc}
\underline{x_{\parallel }}^{\prime } & = &
\begin{array}{c}
\begin{array}{c}
\widetilde{\gamma }(\underline{x_{\parallel }}-\underline{v}t)\medskip
\end{array}
\end{array}
\\
\underline{x_{\perp }}^{\prime } & = &
\begin{array}{c}
\begin{array}{c}
\underline{x_{\perp }}\medskip
\end{array}
\end{array}
\\
t^{\prime } & = & \widetilde{\gamma }\left( t-\sum_{i=1}^{3}\dfrac{%
v^{i}b_{i}^{2}(x^{5})}{c^{2}b_{0}^{2}(x^{5})}x^{i}\right) =\widetilde{\gamma
}(t-\widetilde{\underline{B}}\cdot \underline{x})=\widetilde{\gamma }\left(
t-\widetilde{\underline{B}}^{(\ast )}\ast \underline{x}\right) \smallskip
\end{array}
\right.
\end{equation}
where we put (cfr. Eq.s (22)-(24))
\begin{gather}
\widetilde{\gamma }\equiv (1-\widetilde{\underline{\beta }}\cdot \underline{%
\widetilde{\beta }})^{-1/2}=(1-\widetilde{\underline{\beta }}^{(\ast )}\ast
\underline{\widetilde{\beta }}^{(\ast )})^{-1/2}=  \nonumber \\
\nonumber \\
=\left[ 1-\left( \frac{v^{1}b_{1}(x^{5})}{cb_{0}(x^{5})}\right) ^{2}-\left(
\frac{v^{2}b_{2}(x^{5})}{cb_{0}(x^{5})}\right) ^{2}-\left( \frac{%
v^{3}b_{3}(x^{5})}{cb_{0}(x^{5})}\right) ^{2}\right] ^{-1/2},
\end{gather}

\begin{equation}
\widetilde{\underline{\beta }}^{(\ast )}\equiv \underline{\left( \frac{v}{w}%
\right) }=\left( \frac{v^{1}}{cb_{0}(x^{5})}\widehat{x},\frac{v^{2}}{%
cb_{0}(x^{5})}\widehat{y},\frac{v^{3}}{cb_{0}(x^{5})}\widehat{z}\right) =%
\frac{1}{cb_{0}(x^{5})}\underline{v}\text{ },\text{\ }
\end{equation}
\begin{equation}
\underline{w}\equiv cb_{0}(x^{5})(\widehat{x},\widehat{y},\widehat{z}),\text{
\ }
\end{equation}
\begin{equation}
\widetilde{\underline{B}}\equiv \underline{\left( \frac{v}{u^{2}}\right) }%
=\left( \frac{v^{1}b_{1}^{2}(x^{5})}{c^{2}b_{0}^{2}(x^{5})}\widehat{x},\frac{%
v^{2}b_{2}^{2}(x^{5})}{c^{2}b_{0}^{2}(x^{5})}\widehat{y},\frac{%
v^{3}b_{3}^{2}(x^{5})}{c^{2}b_{0}^{2}(x^{5})}\widehat{z}\right) ,
\end{equation}
\begin{equation}
\widetilde{\underline{B}}^{(\ast )}\equiv \underline{\left( \frac{v}{w^{2}}%
\right) }=\frac{1}{c^{2}b_{0}^{2}(x^{5})}\underline{v}.\text{ \ \ }
\end{equation}

We can therefore state that the deformed boosts admit a {\em double treatment%
}, either:

I) In terms of the Euclidean scalar product $\cdot $, of the (anisotropic)
m.c.v. $\underline{u}$ and of the related ''rapidities'' $\widetilde{%
\underline{\beta }}$ and $\widetilde{\underline{B}}$ ,or

II) in terms of the deformed product $\ast $, of the (isotropic) m.c.v.{\bf %
\ }$\underline{w}$ and of the related quantities $\widetilde{\underline{%
\beta }}^{(\ast )}$ and $\widetilde{\underline{B}}^{(\ast )}$ \footnote{%
It is possible to show that, in this case, more equivalent forms of the
deformed boost (32) exist. As is easily seen, this is due to the fact that,
in general, $\widehat{\widetilde{\beta }}\neq \widehat{v}$ and $\widehat{%
\widetilde{B}}\neq \widehat{v}$ , whereas $\widehat{\widetilde{\beta }%
^{(\ast )}}=\widehat{v}$ $=\widehat{\widetilde{B}^{(\ast )}}$ (cfr. Eq.s
(22), (34), (36) and (37)).}.

Then, the space vector transforms as :
\begin{gather}
\underline{x}^{\prime }=\underline{x_{\parallel }}^{\prime }+\underline{%
x_{\perp }}^{\prime }=\widetilde{\gamma }(\underline{x_{\parallel }}-%
\underline{v}t)+\underline{x_{\perp }}=  \nonumber \\
\nonumber \\
=\underline{x}+(\widetilde{\gamma }-1)\widehat{v}(\widehat{v}\ast \underline{%
x})-\widetilde{\gamma }\underline{v}t=\underline{x}+(\widetilde{\gamma }-1)%
\frac{\underline{v}}{\left| \underline{v}\right| _{\ast }^{2}}(\underline{v}%
\ast \underline{x})-\widetilde{\gamma }\underline{v}t
\end{gather}
and we eventually find the expression of the deformed boost in a generic
direction :
\begin{equation}
\left\{
\begin{array}{ccc}
\underline{x}^{\prime } & = &
\begin{array}{c}
\begin{array}{c}
\underline{x}+(\widetilde{\gamma }-1)\dfrac{\underline{v}}{\left| \underline{%
v}\right| _{\ast }^{2}}(\underline{v}\ast \underline{x})-\widetilde{\gamma }%
\underline{v}t\medskip
\end{array}
\end{array}
\\
t^{\prime } & = & \widetilde{\gamma }\left( t-\widetilde{\underline{B}}\cdot
\underline{x}\right) =\widetilde{\gamma }\left( t-\widetilde{\underline{B}}%
^{(\ast )}\ast \underline{x}\smallskip \right)
\end{array}
\right.
\end{equation}

\

\subsection{Symmetrization of deformed boosts}

As in the case of standard SR, it is possible to symmetrize the expression
of boosts in DSR4 by introducing suitable time coordinates.

Let us first consider a deformed boost along $\widehat{x^{i}}$ ( $i=1,2,3$);
the symmetrization transformation (a {\em ''dimensionally homogenizing
dilato-contration''}) of $t$ is given by

\begin{equation}
x^{0}\equiv u^{i}t=c\frac{b_{0}(x^{5})}{b_{i}(x^{5})}t;\text{ \ }\left(
x^{\prime }\right) ^{i}=\left( x^{i}\right) ^{\prime }\equiv x^{i}
\end{equation}
The deformed metric tensor in the new ''primed'' coordinates,$\left\{ \left(
x^{^{\prime }}\right) ^{\mu }=\left( x^{\mu }\right) ^{\prime }\right\}
=\left\{ x^{0},x,y,z\right\} $, reads :
\begin{gather}
\eta _{\mu \nu }^{\prime }(x^{5})\stackrel{\text{{\footnotesize ESC on}%
{\small \ }}}{=}\eta _{\alpha \beta }(x^{5})\frac{\partial x^{\alpha }}{%
\partial x^{\prime \mu }}\frac{\partial x^{\beta }}{\partial x^{\prime \nu }}%
=diag(b_{i}^{2}(x^{5}),-b_{1}^{2}(x^{5}),-b_{2}^{2}(x^{5}),-b_{3}^{2}(x^{5}))
\nonumber \\
\nonumber \\
\stackrel{\text{{\footnotesize ESC off}{\small \ }}}{=}\delta _{\mu \nu }%
\left[ b_{i}^{2}(x^{5})\delta _{\mu 0}-b_{1}^{2}(x^{5})\delta _{\mu
1}-b_{2}^{2}(x^{5})\delta _{\mu 2}-b_{3}^{2}(x^{5})\delta _{\mu 3}\right]
\end{gather}

Eq. (19) takes therefore the symmetric form\ in $x^{i}$ e $x^{0}$ (ESC off):

\begin{equation}
\left\{
\begin{array}{ccc}
\left( x^{\prime }\right) ^{i}=\left( x^{i}\right) ^{\prime } & = &
\begin{array}{c}
\begin{array}{c}
\widetilde{\gamma }(x^{i}-\widetilde{\beta }^{i}x^{0})\smallskip
\end{array}
\end{array}
\\
\left( x^{\prime }\right) ^{k\neq i}=\left( x^{k\neq i}\right) ^{\prime } & =
&
\begin{array}{c}
\begin{array}{c}
\begin{array}{c}
x^{k\neq i}\bigskip
\end{array}
\end{array}
\end{array}
\\
\left( x^{\prime }\right) ^{0}=\left( x^{0}\right) ^{\prime } & = &
\widetilde{\gamma }(x^{0}-\widetilde{\beta }^{i}x^{i})\smallskip
\end{array}
\right.
\end{equation}
\ \ \

Transformation (40) {\em does not symmetrize }the deformed boost in a
generic direction (unlike the case of SR, where the same transformation $%
x^{0}=ct$ symmetrizes both boosts). In this case, the symmetrization is
possible only \ if the treatment II (based on the deformed scalar product $%
\ast $) is used.

\bigskip In fact, by using the proportionality (already stressed in footnote
8) among $\widetilde{\underline{\beta }}^{(\ast )}$ , $\widetilde{\underline{%
B}}^{(\ast )}$ and $\underline{v}$, the following transformation on $t$

\begin{equation}
x^{0}\equiv cb_{0}(x^{5})t=w^{k}t\text{ \ }(\forall k=1,2,3)\text{\ \ ; }%
\left( x^{\prime }\right) ^{i}=\left( x^{i}\right) ^{\prime }\equiv x^{i}%
\text{ \ \ \ \ \ }(\forall i=1,2,3)
\end{equation}
does symmetrize Eq. (32) in $\underline{x_{\parallel }}$ e $x^{0}$:
\begin{equation}
\left\{
\begin{array}{ccc}
\underline{x_{\parallel }}^{\prime } & = &
\begin{array}{c}
\begin{array}{c}
(1-\widetilde{\underline{\beta }}^{(\ast )}\ast \underline{\widetilde{\beta }%
}^{(\ast )})^{-1/2}(\underline{x_{\parallel }}-\underline{\widetilde{\beta }}%
^{(\ast )}x^{0})\bigskip
\end{array}
\end{array}
\\
\underline{x_{\perp }}^{\prime } & = &
\begin{array}{c}
\begin{array}{c}
\begin{array}{c}
\underline{x_{\perp }}\bigskip
\end{array}
\end{array}
\end{array}
\\
\left( x^{\prime }\right) ^{0}=\left( x^{0}\right) ^{\prime } & = & \left\{
\begin{array}{c}
(1-\widetilde{\underline{\beta }}^{(\ast )}\ast \underline{\widetilde{\beta }%
}^{(\ast )})^{-1/2}(x^{0}-\widetilde{\underline{\beta }}^{(\ast )}\ast
\underline{x})\text{ }=\smallskip \\
\\
=(1-\widetilde{\underline{\beta }}^{(\ast )}\ast \underline{\widetilde{\beta
}}^{(\ast )})^{-1/2}(x^{0}-\widetilde{\underline{\beta }}^{(\ast )}\ast
\underline{x_{\parallel }})\smallskip
\end{array}
\right.
\end{array}
\right.
\end{equation}

Under transformation (43), the metric tensor becomes :
\begin{gather}
\eta _{\mu \nu }^{\prime }(x^{5})\stackrel{\text{{\footnotesize ESC on}%
{\small \ }}}{=}\eta _{\alpha \beta }(x^{5})\frac{\partial x^{\alpha }}{%
\partial x^{\prime \mu }}\frac{\partial x^{\beta }}{\partial x^{\prime \nu }}%
=diag(1,-b_{1}^{2}(x^{5}),-b_{2}^{2}(x^{5}),-b_{3}^{2}(x^{5}))  \nonumber \\
\nonumber \\
\stackrel{\text{{\footnotesize ESC off}{\small \ }}}{=}\delta _{\mu \nu }%
\left[ \delta _{\mu 0}-b_{1}^{2}(x^{5})\delta _{\mu
1}-b_{2}^{2}(x^{5})\delta _{\mu 2}-b_{3}^{2}(x^{5})\delta _{\mu 3}\right]
\end{gather}
Therefore the symmetrization of the deformed boost in a generic direction
makes the 4-d. metric {\em isochronous}, since $\eta _{00}^{\prime }=1$ so
that $\tau =t$ (namely {\em proper time} coincides with {\em coordinate time}%
) .

Let us finally notice that, like in the SR case, the boost in generic
direction expressed in terms of $\underline{x}$ e $t$ (Eq. (39)) {\em cannot}
in general be symmetrized.

\bigskip

\subsection{Velocity composition law in $\widetilde{M_{4}}$ \ and the
invariant maximal speed}

\bigskip

We have seen in Subsect. 2.2 that the directionally separating approach
(mandatory in the deformed case) yields two different {\em mathematical}
definitions $\underline{u}$ (Eq. (14)) and $\underline{w}$\ \ (Eq. (16))$\ \
$of maximal causal velocity in DSR4. The choice between them must be done on
a physical basis, by checking their actual invariance under deformed boosts.

To this aim, we have to derive the generalized velocity composition law
valid in $\widetilde{M_{4}}$ . For a deformed boost in the direction $%
\widehat{x^{i}}$ , we have, by differentiating the inverse of Eq. (19) (on
account of the fact that $dx^{5}=0$ in DSR4) (ESC off):
\begin{equation}
\left\{
\begin{array}{ccc}
dx^{i} & = &
\begin{array}{c}
\begin{array}{c}
\widetilde{\gamma }\left[ \left( dx^{i}\right) ^{\prime }+v^{i}\left(
dt\right) ^{\prime }\right] \ \smallskip
\end{array}
\end{array}
\\
dx^{k\neq i}\smallskip \smallskip & = &
\begin{array}{c}
\begin{array}{c}
\left( dx^{k\neq i}\right) ^{\prime }\smallskip \smallskip
\end{array}
\end{array}
\\
dt & = & \widetilde{\gamma }\left[ \left( dt\right) ^{\prime }+\dfrac{%
v^{i}b_{i}^{2}(x^{5})}{c^{2}b_{0}^{2}(x^{5})}\left( dx^{i}\right) ^{\prime }%
\right]
\end{array}
\right.
\end{equation}
with $\widetilde{\gamma }$ given by (21). Since
\begin{equation}
\left\{
\begin{array}{l}
\frac{dx^{i}}{dt}=v^{i}, \\
\\
\frac{\left( dx^{i}\right) ^{\prime }}{\left( dt\right) ^{\prime }}=\frac{%
\left( dx^{^{\prime }}\right) ^{i}}{\left( dt\right) ^{\prime }}=\left(
v^{\prime }\right) ^{i}=\left( v^{i}\right) ^{\prime }, \\
\\
\frac{dx^{k\neq i}}{dt}=v^{k\neq i}, \\
\\
\frac{\left( dx^{k\neq i}\right) ^{\prime }}{\left( dt\right) ^{\prime }}=%
\frac{\left( dx^{^{\prime }}\right) ^{k\neq i}}{\left( dt\right) ^{\prime }}%
=\left( v^{\prime }\right) ^{k\neq i}=\left( v^{k\neq i}\right) ^{\prime },
\end{array}
\right.
\end{equation}
we get the {\em deformed velocity composition law }(in compact notation, ESC
off)
\begin{equation}
v^{k}=\frac{\left( v^{k}\right) ^{\prime }+\delta _{ik}v^{i}}{\left[
1+\left( \dfrac{b_{i}(x^{5})}{b_{0}(x^{5})}\right) ^{2}\dfrac{v^{i}\left(
v^{i}\right) ^{\prime }}{c^{2}}\right] \left\{ \tilde{\gamma}(x^{5})+\delta
_{ik}\left[ 1-\tilde{\gamma}(x^{5})\right] \right\} }
\end{equation}

This relation can be expressed in terms of the standard 3-d. scalar product $%
\cdot $ (and therefore of the anisotropic maximal velocity $\underline{u}$)
(approach I) as
\begin{gather}
v^{k}=\frac{\left( v^{k}\right) ^{\prime }+\delta _{ik}v^{i}}{\left[ 1+\
\dfrac{\underline{v}\cdot \underline{v}^{\prime }}{\left(
u^{i}(x^{5})\right) ^{2}}\right] \left\{ \tilde{\gamma}(x^{5})+\delta _{ik}%
\left[ 1-\tilde{\gamma}(x^{5})\right] \right\} }=  \nonumber \\
\nonumber \\
=\frac{\left( v^{k}\right) ^{\prime }+\delta _{ik}v^{i}}{\left[ 1+\ \dfrac{%
\underline{\tilde{\beta}}\cdot \underline{v}^{\prime }}{u^{i}(x^{5})}\right]
\left\{ \tilde{\gamma}(x^{5})+\delta _{ik}\left[ 1-\tilde{\gamma}(x^{5})%
\right] \right\} }
\end{gather}
where (cfr. Eq.s (22),(23))
\begin{equation}
\tilde{\beta}^{i}(x^{5})=\frac{v^{i}}{u^{i}(x^{5})}\text{ \ ; \ }\tilde{%
\gamma}(x^{5})=\left( 1-\underline{\tilde{\beta}}(x^{5})\cdot \underline{%
\tilde{\beta}}(x^{5})\right) ^{-1/2}
\end{equation}
Alternatively, we can use approach II, based on the deformed scalar product $%
\ast $ (and therefore the isotropic maximal velocity $\underline{w}$) and
write Eq.(48) as
\begin{gather}
v^{k}=\frac{\left( v^{k}\right) ^{\prime }+\delta _{ik}v^{i}}{\left[ 1+\
\dfrac{\underline{v}\ast \underline{v}^{\prime }}{\left( w^{i}(x^{5})\right)
^{2}}\right] \left\{ \tilde{\gamma}(x^{5})+\delta _{ik}\left[ 1-\tilde{\gamma%
}(x^{5})\right] \right\} }=  \nonumber \\
\nonumber \\
=\frac{\left( v^{k}\right) ^{\prime }+\delta _{ik}v^{i}}{\left[ 1+\ \dfrac{%
\underline{\tilde{\beta}^{(\ast )}}\ast \underline{v}^{\prime }}{w^{i}(x^{5})%
}\right] \left\{ \tilde{\gamma}(x^{5})+\delta _{ik}\left[ 1-\tilde{\gamma}%
(x^{5})\right] \right\} }
\end{gather}
with (cfr. Eq.s (34), (33))
\begin{equation}
\tilde{\beta}^{(\ast )i}(x^{5})=\frac{v^{i}}{w^{i}(x^{5})}\text{ \ ; \ }%
\tilde{\gamma}(x^{5})=\left( 1-\underline{\tilde{\beta}}^{\ast }(x^{5})\ast
\underline{\tilde{\beta}}^{\ast }(x^{5})\right) ^{-1/2}
\end{equation}

It is now an easy task to check the truly maximal character of the two
velocities. Indeed, if $\left( v^{i}\right) ^{\prime }=u^{i}(x^{5})$, one
gets, from Eq.(49)
\begin{equation}
v^{i}=\frac{u^{i}(x^{5})+v^{i}}{1+\dfrac{v^{i}}{u^{i}(x^{5})}}=u^{i}(x^{5})
\end{equation}
whereas, for $\left( v^{i}\right) ^{\prime }=w^{i}(x^{5})$, Eq. (51) yields
\begin{equation}
v^{i}=\frac{w^{i}(x^{5})+v^{i}}{1+\dfrac{\left( b_{i}(x^{5})\right) ^{2}v^{i}%
}{w^{i}(x^{5})}}\neq w^{i}(x^{5})
\end{equation}

We can therefore conclude, on a physical basis, that {\em \ }$\underline{u}$%
{\em \ is the maximal, invariant causal velocity in DSR4, and it plays in
the deformed Minkowski space }$\widetilde{M_{4}}${\em \ the role of the
light speed in standard SR}\footnote{%
Of course, in the case of space isotropy (cfr. Eq.(4)), we recover the
result of Ref. [1], namely an isotropic maximal causal velocity given by
\begin{gather*}
u_{iso}^{i}(x^{5})=\left. u_{DSR4,II}^{i}(x^{5})\right|
_{b_{i}(x^{5})=b(x^{5})}=c\frac{b_{0}(x^{5})}{b(x^{5})}\text{ \ \ \ \ }%
\forall i=1,2,3, \\
\\
\left| \underline{u}_{iso}(x^{5})\right| =\left( \sum_{i=1}^{3}\left(
u_{iso}^{i}(x^{5})\right) ^{2}\right) ^{1/2}=\sqrt{3}c\frac{b_{0}(x^{5})}{%
b(x^{5})}
\end{gather*}
Moreover, notice that it can be
\[
u^{i}{\gtreqqless }c\,\,\,\text{according\thinspace\ to}\,\,\frac{b_{0}}{%
b^{i}}\gtreqqless 1\ .
\]
\par
In other words, there may be maximal causal speeds {\em superluminal},
depending on the interaction considered (without {\em any} violation of
causality).}.

\bigskip It is also easy to see why - although approach II) looks at first
sight more rigorous mathematically, because it permits to connect the
peculiar features of spatial anisotropy of DSR4 to the deformed product $%
\ast $, ''naturally induced'' from the metric of $\widetilde{M_{4}}(x^{5})$
- actually it is approach I) which yields the physically relevant result.
Indeed, the velocity $\underline{u}$ is just defined as $\dfrac{d\underline{x%
}}{dt}$, and it therefore represents the physically measured velocity, for a
particle moving in the usual, physical Euclidean 3-d space. On the other
hand, this result clearly shows that the space anisotropy introduced by the
deformed metric {\em is not a mere mathematical artifact, but it reflects
itself in the physical properties (imposed by the interaction involved) of \
the phenomenon described by the deformed space-time}$.$

The comparison of the deformed boost expressions (Eq.s (19), (32)) with the
corresponding ones of the standard Lorentz boosts shows clearly that the
transition from SR (based on $M_{4}$) to DSR4 (based on $\widetilde{M_{4}}$)
is simply carried out by letting
\begin{equation}
\underline{u}_{SR}=c(\widehat{x},\widehat{y},\widehat{z})\longrightarrow
\underline{u}_{DSR4}(x^{5})=cb_{0}(x^{5})\left( \frac{1}{b_{1}(x^{5})}%
\widehat{x},\frac{1}{b_{2}(x^{5})}\widehat{y},\frac{1}{b_{3}(x^{5})}\widehat{%
z}\right)
\end{equation}

In other words, the difference between $M_{4}$ and $\widetilde{M_{4}}(x^{5})$
(at least as far as the finite coordinate transformations are concerned) is
completely embodied in the trivector m.c.v. $\underline{u}$.

\subsection{\protect\bigskip Choosing the boost direction in DSR4}

We want now to remark a difficulty arising in the context of DSR4, due to
the space anisotropy.

Indeed, the space anisotropy (reflected in the physical anisotropic m.c.v. $%
\underline{u}$) produces a triple indetermination in the process of
identifying the motion axis with any of the space coordinate axes, since now
- unlike the SR case - the space dimensions are no longer equivalent.

However, this indeterminacy can be removed (at least in principle) by means
of the following {\it \ }{\em Gedankenexperiment} . Consider three particles
(ruled by one and the same interaction) in general different but able to
move at the maximal causal velocity $u^{i}(x^{5})$.\ Suppose they are moving
in the 3-d Euclidean space along mutually independent (orthogonal) spatial
directions. Assigning an arbitrary labelling to the particle motion
directions, we can fix an orthogonal, left-handed fame of axes. Since by
assumption we know the interaction which the particles are subjected to, we
know the deformed metric and therefore the metric coefficients as functions
of the energy, $b_{\mu }^{2}(E)$. Then, a measurement of the particle
velocities allows us to determine the right labelling of the spatial frame
(cfr. Eq. (24)).

This implies that in the context of DSR4, too, it is always possible, at
physical level, to let one of the three space axes to coincide with the
direction of motion of a physical object, and therefore apply the suitable
deformed boost.

\bigskip

\paragraph{Appendix A}

\bigskip

Consider two TIRF's, $K$ and $K^{\prime }$; by definition, the DLT's are
metric isometries, i.e. leave {\em invariant} the deformed metric interval
(2), whence:
\begin{gather}
b_{0}^{2}(x^{5})c^{2}t^{2}-b_{1}^{2}(x^{5})x^{2}-b_{2}^{2}(x^{5})y^{2}-b_{3}^{2}(x^{5})z^{2}=
\nonumber \\
\nonumber \\
=b_{0}^{2}(x^{5})c^{2}\left( t^{\prime }\right) ^{2}-b_{1}^{2}(x^{5})\left(
x^{\prime }\right) ^{2}-b_{2}^{2}(x^{5})\left( y^{\prime }\right)
^{2}-b_{3}^{2}(x^{5})\left( z^{\prime }\right) ^{2}  \tag{A.1}
\end{gather}

Moreover, without loss of generality, we can assume that the frames $K$ and $%
K^{\prime }$ are in {\em standard configuration} (i.e. their spatial frames
coincide at $t=t^{\prime }=0$). By choosing the boost direction along\ $%
\widehat{x^{1}}=\widehat{x}$ , we have therefore $y^{\prime }=y$ , $%
z^{\prime }=z$, and Eq. (A.1) reduces to

\begin{equation}
b_{0}^{2}(x^{5})c^{2}t^{2}-b_{1}^{2}(x^{5})x^{2}=b_{0}^{2}(x^{5})c^{2}\left(
t^{\prime }\right) ^{2}-b_{1}^{2}(x^{5})\left( x^{\prime }\right) ^{2}
\tag{A.2}
\end{equation}
From {\em space-time homogeneity} it follows that the functional relations
between the two sets of coordinates $\left\{ x,y,z,t\right\} $ and $\left\{
x^{\prime },y^{\prime },z^{\prime },t^{\prime }\right\} $ must be {\em linear%
}. Then, in general, the deformed coordinate transformations are to be
searched in the form
\begin{equation}
\left\{
\begin{array}{ccc}
x^{\prime } & = & A_{11}x+A_{14}t \\
y^{\prime } & = & y \\
z^{\prime } & = & z \\
t^{\prime } & = & A_{41}x+A_{44}t
\end{array}
\right.  \tag{A.3}
\end{equation}
where the coefficients $A_{11},A_{14},A_{41},A_{44}$ depend {\it \ a priori}
in general on $\underline{v}$ and $\widehat{x}$ (and, parametrically, on $%
x^{5}$).

Notice that the origin $O^{\prime }$ of TIRF $K^{\prime }$ must move in $K$
with velocity $\underline{v}=v^{1}\widehat{x}$ , and therefore :

\begin{equation}
x^{\prime }=0\text{ \ , \ }x=vt\Leftrightarrow
A_{14}=-vA_{11}\Leftrightarrow x^{\prime }=A_{11}(x-vt)  \tag{A.4}
\end{equation}
Replacing (A.3), (A.4) in (A.2) yields
\begin{equation}
b_{0}^{2}(x^{5})c^{2}t^{2}-b_{1}^{2}(x^{5})x^{2}=b_{0}^{2}(x^{5})c^{2}(A_{41}x+A_{44}t)^{2}-A_{11}^{2}b_{1}^{2}(x^{5})x^{2}(x-vt)^{2}
\tag{A.5}
\end{equation}
which implies the following $3\times 3$ quadratic system:

\begin{equation}
\left\{
\begin{array}{ccc}
c^{2} & = &
\begin{array}{c}
c^{2}A_{44}^{2}-\left( \frac{b_{1}(x^{5})}{b_{0}(x^{5})}\right)
^{2}A_{11}^{2}v^{2}\smallskip
\end{array}
\\
-1 & = &
\begin{array}{c}
c^{2}\left( \frac{b_{0}(x^{5})}{b_{1}(x^{5})}\right)
^{2}A_{41}^{2}-A_{11}^{2}\smallskip
\end{array}
\\
0 & = & c^{2}\left( \frac{b_{0}(x^{5})}{b_{1}(x^{5})}\right)
^{2}A_{41}A_{44}+A_{11}^{2}v\smallskip
\end{array}
\right.  \tag{A.6}
\end{equation}
with general solution
\begin{equation}
A_{11}=A_{44}=\pm \left( 1-\left( \frac{vb_{1}(x^{5})}{cb_{0}(x^{5})}\right)
^{2}\right) ^{-1/2}\ \ \   \tag{A.7}
\end{equation}
\begin{equation}
\ \ A_{14}=\mp \left( \frac{vb_{1}^{2}(x^{5})}{c^{2}b_{0}^{2}(x^{5})}\right)
\left( 1-\left( \frac{vb_{1}(x^{5})}{cb_{0}(x^{5})}\right) ^{2}\right)
^{-1/2}=-\left( \frac{vb_{1}^{2}(x^{5})}{c^{2}b_{0}^{2}(x^{5})}\right) A_{11}
\tag{A.7'}
\end{equation}
The final result is

\begin{equation}
\left\{
\begin{array}{ccc}
x^{\prime } & = &
\begin{array}{c}
\widetilde{\gamma }(x-vt)=\ \ \widetilde{\gamma }\left( x-\widetilde{\beta }%
\dfrac{b_{0}(x^{5})}{b_{1}(x^{5})}ct\right)
\end{array}
\\
y^{\prime } & = &
\begin{array}{c}
y
\end{array}
\\
z^{\prime } & = &
\begin{array}{c}
z
\end{array}
\\
t^{\prime } & = & \widetilde{\gamma }\left( t-\dfrac{vb_{1}^{2}(x^{5})}{%
c^{2}b_{0}^{2}(x^{5})}x\right) =\widetilde{\gamma }\left( t-\dfrac{%
\widetilde{\beta }^{2}}{v}x\right)
\end{array}
\right.  \tag{A.8}
\end{equation}
namely the deformed boost (19) for motion along $x$.

The same procedure can in principle be followed in deriving the deformed
boost in a generic direction. In this case, the coordinate transformations
are

\begin{equation}
\left\{
\begin{array}{ccc}
x^{\prime } & = &
\begin{array}{c}
A_{11}x+A_{12}y+A_{13}z+A_{14}t\smallskip
\end{array}
\\
y^{\prime } & = &
\begin{array}{c}
A_{21}x+A_{22}y+A_{23}z+A_{24}t\smallskip
\end{array}
\\
z^{\prime } & = &
\begin{array}{c}
A_{31}x+A_{32}y+A_{33}z+A_{34}t\smallskip
\end{array}
\\
t^{\prime } & = & A_{41}x+A_{42}y+A_{43}z+A_{44}t\smallskip
\end{array}
\right.  \tag{A.9}
\end{equation}
From the physical requirement that the origin $O^{\prime }$ of TIRF $%
K^{\prime }$ must move in $K$ with velocity components $v^{1}$ along $%
\widehat{x}$, $v^{2}$ along $\widehat{y}$, $v^{3}$ along $\widehat{z}$ , one
gets:

\begin{gather}
\left.
\begin{array}{ccccccc}
x^{\prime } & = & 0 & , & x & = & v^{1}t \\
y^{\prime } & = & 0 & , & y & = & v^{2}t \\
z^{\prime } & = & 0 & , & z & = & v^{3}t
\end{array}
\right\} \Leftrightarrow  \nonumber \\
\nonumber \\
\nonumber \\
\Leftrightarrow \left\{
\begin{array}{ccccccccc}
A_{11}v^{1} & + & A_{12}v^{2} & + & A_{13}v^{3} & + & A_{14} & = & 0 \\
A_{21}v^{1} & + & A_{21}v^{2} & + & A_{31}v^{3} & + & A_{24} & = & 0 \\
A_{31}v^{1} & + & A_{32}v^{2} & + & A_{33}v^{3} & + & A_{34} & = & 0
\end{array}
\right.  \tag{A.10}
\end{gather}
Eq. (A.9) becomes therefore

\begin{equation}
\left\{
\begin{array}{ccc}
x^{\prime } & = &
\begin{array}{c}
A_{11}(x-v^{1}t)+A_{12}(y-v^{2}t)+A_{13}(z-v^{3}t)\smallskip
\end{array}
\\
y^{\prime } & = &
\begin{array}{c}
A_{21}(x-v^{1}t)+A_{22}(y-v^{2}t)+A_{23}(z-v^{3}t)\smallskip
\end{array}
\\
z^{\prime } & = &
\begin{array}{c}
A_{31}(x-v^{1}t)+A_{32}(y-v^{2}t)+A_{33}(z-v^{3}t)\smallskip
\end{array}
\\
t^{\prime } & = & A_{41}x+A_{42}y+A_{43}z+A_{44}t\smallskip
\end{array}
\right.  \tag{A.11}
\end{equation}

Replacing (A.11) in (A.1) yields

\begin{gather}
b_{0}^{2}(x^{5})c^{2}t^{2}-b_{1}^{2}(x^{5})x^{2}-b_{2}^{2}(x^{5})y^{2}-b_{3}^{2}(x^{5})z^{2}=
\nonumber \\
\smallskip  \nonumber \\
\nonumber \\
=c^{2}b_{0}^{2}(x^{5})(A_{41}x+A_{42}y+A_{43}z+A_{44}t)^{2}+  \nonumber \\
\nonumber \\
-b_{1}^{2}(x^{5})(A_{11}(x-v^{1}t)+A_{12}(y-v^{2}t)+A_{13}(z-v^{3}t))^{2}+
\nonumber \\
\nonumber \\
-b_{2}^{2}(x^{5})(A_{21}(x-v^{1}t)+A_{22}(y-v^{2}t)+A_{23}(z-v^{3}t))^{2}+
\nonumber \\
\nonumber \\
-b_{3}^{2}(x^{5})(A_{31}(x-v^{1}t)+A_{32}(y-v^{2}t)+A_{33}(z-v^{3}t))^{2}
\tag{A.12}
\end{gather}
Equating the coefficients on both sides of (A.12) one gets a system of 10
quadratic equations in the 13 unknown coefficients $\left\{
A_{ij},A_{4i}\right\} $ $(i,j=1,2,3)$, namely :

I- from the coefficient of $t^{2}$ :
\begin{equation}
c^{2}(A_{44}^{2}-1)-\frac{1}{b_{0}^{2}(x^{5})}%
\sum_{i,j,l=1}^{3}b_{j}^{2}(x^{5})v^{j}v^{l}A_{ij}A_{il}=0  \tag{A.13}
\end{equation}

II- from the coefficients of $x^{i}x^{j}$ ($i,j=1,2,3$), 6 independent
equations:
\begin{equation}
c^{2}A_{4i}A_{4j}-\frac{1}{b_{0}^{2}(x^{5})}%
\sum_{l=1}^{3}b_{l}^{2}(x^{5})(A_{li}A_{lj}-\delta _{ij}\delta _{il})=0
\tag{A.14}
\end{equation}

III- from the coefficients of $x^{i}t$ ($i=1,2,3$), 3 independent equations:
\begin{equation}
c^{2}A_{4i}A_{44}+\frac{1}{b_{0}^{2}(x^{5})}%
\sum_{j,l=1}^{3}b_{j}^{2}(x^{5})v^{l}A_{ji}A_{jl}=0  \tag{A.15}
\end{equation}
Although the above system in the set $\left\{ A_{ij},A_{4i}\right\} $ $%
(i,j=1,2,3)$ can be exactly solved, the general solution for the boost
expressed in the form (A.11) is quite cumbersome. This motivates our choice
(adopted in Subsect. 3.2) of deriving the form of the deformed boost in a
generic direction by exploiting the notion of ''deformed'' parallelism
between trivectors. Of course (and it may be checked by explicit
calculations), both approaches are completely {\em equivalent}.

\ \ \ \ \ \ \ \ \ \ \ \ \ \ \ \ \ \ \ \ \ \ \ \ \ \ \ \ \ \ \ \ \ \ \ \ \ \
\ \ \ \ \ \ \ \ \ \ \ \ \ \ \ \ \ \ \ \ \ \ \ \ \ \ \ \ \ \ \ \ \ \ \ \ \ \
\ \

\bigskip {\bf References\bigskip }

\bigskip {\bf \lbrack 1]} \ F. Cardone and R. Mignani: ''On a nonlocal
relativistic kinematics'', INFN\medskip\ preprint n.910 (Roma, Nov. 1992);
{\it Grav. \& Cosm.} {\bf 4} , 311 (1998);\ {\it Found. Phys. }{\bf 29},
1735 (1999); \ {\it Ann. Fond. L. de Broglie }{\bf 25 }, 165 (2000).

\bigskip

{\bf [2]} \ F. Cardone, R. Mignani, and R.M. Santilli: {\it J. Phys.}G {\bf %
18}, L61, L141 (1992).\medskip

{\bf [3]} \ F. Cardone and R. Mignani: \ {\it JETP} {\bf 83}, 435 [{\it Zh.
Eksp. Teor. Fiz.} {\bf 110}, 793]\medskip\ (1996); F. Cardone, M. Gaspero,
and R. Mignani: {\it Eur. Phys. J.} C {\bf 4}, 705 (1998).\medskip

{\bf [4]} F.Cardone e R.Mignani : {\it Ann. Fond. L. de Broglie}, {\bf 23} ,
173 (1998);\ F. Cardone, R. Mignani, and V.S. Olkhovski: {\it J. de Phys.I
(France)}{\bf \ 7}, 1211\medskip\ (1997); {\it Modern Phys. Lett.} B {\bf 14}%
, 109 (2000).\medskip

{\bf [5]} \ F. Cardone and R. Mignani: {\it Int. J. Modern Phys.} A {\bf 14}%
, 3799 (1999).\medskip

{\bf [6]} \ F. Cardone, M. Francaviglia, and R. Mignani: {\it Gen. Rel. Grav.%
} {\bf 30}, 1619\medskip\ (1998); {\it ibidem}, {\bf 31}, 1049 (1999); {\it %
Found. Phys.\medskip\ Lett.} {\bf 12}, 281, 347 (1999){\it .\medskip }

{\bf [7] }A. Marrani: {\it ''Simmetrie di Killing di Spazi di Minkowski
generalizzati''} ({\it ''Killing Symmetries of Generalized Minkowski Spaces''%
}) (Laurea Thesis), Rome, October 2001 (in Italian).

\end{document}